\pdfoutput=1
\documentclass[10pt,conference]{IEEEtran}
\usepackage{cite}
\usepackage{amsmath,amssymb,amsfonts}
\usepackage{graphicx}
\usepackage{textcomp}
\usepackage{xcolor}
\usepackage[hyphens]{url}
\usepackage{fancyhdr}
\usepackage{hyperref}
\usepackage{booktabs}
\usepackage{multirow}
\usepackage{pifont}
\usepackage[ruled,vlined,linesnumbered]{algorithm2e}
\SetNlSty{normalfont}{}{}
\SetKwInput{KwInput}{Input}
\SetKwInput{KwOutput}{Output}
\SetKwComment{Comment}{// }{}

\SetCommentSty{algcommentfont}
\SetAlgoNlRelativeSize{0}
\usepackage{array}

\definecolor{revblue}{RGB}{0,90,220}
\definecolor{revred}{RGB}{205,55,55}

\usepackage{colortbl}  
\usepackage{tabularx}
\usepackage{siunitx}
\usepackage{nicematrix} 

\title{A Differential Computing Accelerator for LLM Inference Leveraging Bit-Slice Transition Sparsity}

\newcommand{\authorinfo}[4]{%
  \parbox[t]{0.31\textwidth}{%
    \centering
    {\normalfont\fontsize{12}{13}\selectfont #1}\\[-1pt]
    {\normalfont\fontsize{10}{12}\selectfont
      #2\\
      #3\\[-2pt]
      #4\par
    }
  }%
}

\newcommand{\ustc}{
  University of Science and\\
  Technology of China
}

\newcommand{\hpcaauthors}{%
  \begin{tabular}{@{}c@{}}
    \authorinfo{Yifan Wang}
      {\ustc}
      {Hefei, China}
      {wangyif513@mail.ustc.edu.cn}
    \hspace{0.015\textwidth}
    \authorinfo{Zetao Guo}
      {\ustc}
      {Hefei, China}
      {ztguo@mail.ustc.edu.cn}
    \hspace{0.015\textwidth}
    \authorinfo{You Xiao}
      {\ustc}
      {Hefei, China}
      {youxiao@mail.ustc.edu.cn}
    \\ \noalign{\vskip 12pt}
    \authorinfo{Wendi Sun}
      {\ustc}
      {Hefei, China}
      {wdsun@mail.ustc.edu.cn}
    \hspace{0.03\textwidth}
    \authorinfo{Song Chen}
      {\ustc}
      {Hefei, China}
      {songch@ustc.edu.cn}
  \end{tabular}
}

\author{\hpcaauthors}

\begin{document}
\maketitle
\thispagestyle{plain}
\pagestyle{plain}


\begin{abstract}

Large language models (LLMs) have become foundational to modern artificial intelligence. Driven by scaling laws, their rapid growth poses increasing computational and memory challenges for deployment. To accelerate the general matrix multiplication (GEMM) operations that dominate LLM computation, bit-slice architectures decompose quantized operands along the bit dimension, exposing slice-level sparsity and redundancy. However, existing bit-slice designs primarily exploit sparsity in high-order slices, while dense low-order slices still offer substantial untapped reuse opportunities.

We observe that bit-slice vectors (BSVs) are highly clustered in Hamming space, with many BSVs lying only a few bit flips away from a small set of representative patterns. Based on this observation, we propose ISTS, an algorithm-hardware co-designed accelerator that leverages representative BSV patterns as shared Steiner points to reuse partial sums across slices, substantially reducing computation. Specifically, the partial sum of each target BSV can be reconstructed from that of its assigned Steiner point using only a few additions or subtractions. On the algorithmic side, we propose Steiner Absorption Table Construction (SATC), which selects a compact set of Steiner points to absorb BSVs with only a few bidirectional bit flips, while FlipPrune selectively prunes flips to further expand the reusable BSV population. On the hardware side, we design a lightweight accelerator that supports Steiner reuse for static linear layer weights and runtime-generated attention K/V tensors. Comprehensive evaluations show that ISTS achieves average speedups of $3.09\times$, $2.41\times$, $2.18\times$, $1.27\times$, and $1.41\times$ over the state-of-the-art accelerators ANT, BBS, LUT Tensor Core, Transitive Array, and MCBP-BRCR, respectively.

\end{abstract}

\section{Introduction}
Large language models (LLMs) have demonstrated remarkable capabilities across a wide range of applications, including content generation~\cite{gpt3,instructgpt}, software development~\cite{codex,alphacode}, healthcare~\cite{medpalm}, scientific discovery~\cite{coscientist,chemcrow}, and embodied robotics~\cite{palme,rt2}. As model sizes and context lengths continue to grow, LLM inference demands greater computational throughput, higher energy efficiency, and more memory bandwidth. General matrix multiplication (GEMM) is a dominant computational kernel in LLM inference, including the large matrix operations in linear projections, attention computation, and feed-forward networks throughout each Transformer layer\cite{lin2025qserve}\cite{zhao2025ganq}. Low-bit quantization has therefore become a mainstream optimization for LLM deployment, reducing model storage and memory traffic while enabling lower-cost arithmetic through reduced data bit widths~\cite{Atom,awq,efficientqat}.

\begin{figure}[h]
    \centering
    \includegraphics[width=1\linewidth]{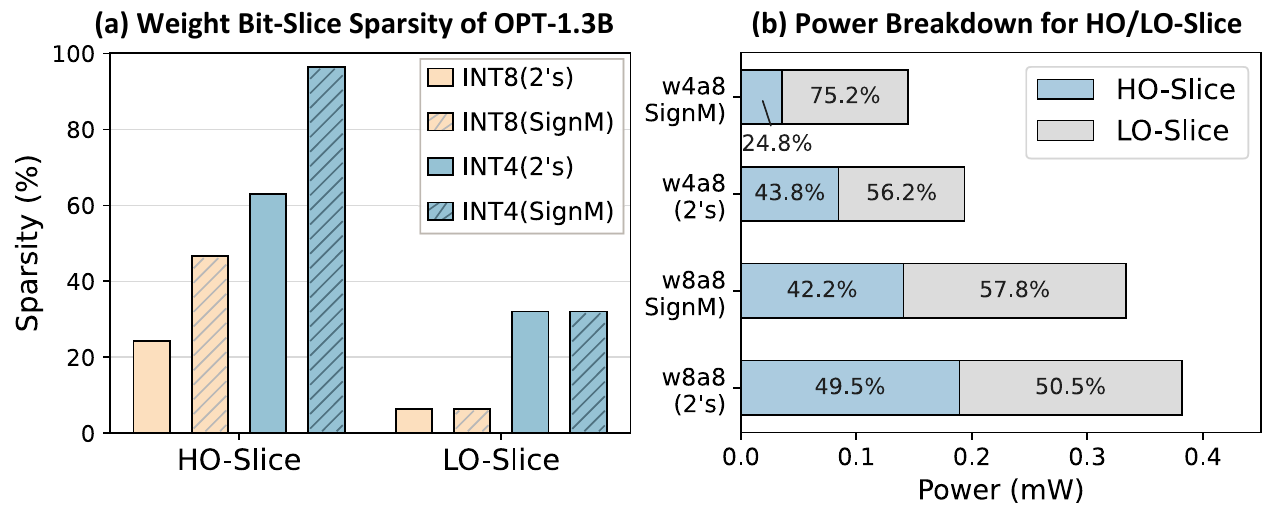}
    \caption{Bit-slice sparsity and PE power distribution under different weight precisions and signed representations. (a) Average sparsity of the HO/LO weight slices under two's-complement (2's) and sign-magnitude (SignM) representations. For SignM, the sign bit is handled separately and excluded from the HO-slice. (b) Power breakdown under W8A8 and W4A8.}
    \label{fig1}
\end{figure}

Bit-slice architectures accelerate quantized DNN execution by exploiting slice-level sparsity at zero and near-zero data~\cite{Bit-slice-Architecture,han2021hnpu,sibia}. They use low-precision slices as the basic execution granularity, decomposing each integer operand into multiple slices, such as partitioning an 8-bit value into two 4-bit slices, and skipping ineffectual slices during execution. In conventional DNNs such as ResNet~\cite{resnet}, weights are concentrated around zero, while ReLU sets negative activations to zero, causing high-order (HO) slices to often exceed 70\% sparsity~\cite{Lutein}. However, this advantage diminishes in low-bit LLMs, where GELU and SiLU preserve negative responses and therefore produce denser activations~\cite{song2024prosparse,hendrycks2016gelu,shazeer2020glu}. HO-slice sparsity is also strongly affected by the signed representation. As shown in Fig.~\ref{fig1}(a), sign-magnitude representation yields substantially sparser HO-slices than two's complement, but its HO region contains only one skippable bit under 4-bit quantization. Consequently, dense low-order (LO) slice processing dominates the remaining computation and accounts for 50.5\%--70.2\% of the total power across different quantization configurations, as shown in Fig.~\ref{fig1}(b). Therefore, accelerating low-bit LLMs requires not only exploiting sparse MSBs but also uncovering bit-level reuse opportunities within dense and irregular LO-slices.

Recent work moves beyond zero skipping by exploiting reuse to reduce bit-slice computation. MCBP~\cite{MCBP} identifies repeated bit-column patterns and merges their corresponding activation terms, so each shared partial sum (psum) is computed only once. Transitive Array (TA)~\cite{Transitive-Array} introduces transitive sparsity to reuse previously computed TransRow partial sums. For example, the partial sum of $\mathtt{0111}$ can be derived from that of $\mathtt{0011}$ by computing only the contribution of the newly introduced 1-bit.
While these designs demonstrate the effectiveness of bit-slice reuse, several limitations remain. MCBP requires exact repetition, leaving near-matching columns with even a 1-bit difference unable to share computation. TA's monotonic reuse relation induces a partial order among TransRows, naturally represented as a Hasse graph. Organizing these relations into reuse trees introduces sequential dependencies along each path and may cause workload imbalance across trees, requiring scoreboard based scheduling and balanced forest mapping, which increase control complexity and scheduling overhead. 

We observe that many BSVs exhibit low Hamming-distance transitions. Motivated by the above limitations, we propose ISTS, which organizes BSVs around a small set of shared Steiner points. Each Steiner point enables parallel reconstruction of multiple nearby BSVs with only a few add/subtract operations. As illustrated in Fig.~\ref{fig3}~(a), two Steiner points cover six BSVs. We further design an efficient accelerator that integrates Steiner-point construction, transition encoding, and hardware execution. Our main contributions are summarized as follows:

\begin{itemize}
\item We formulate a BSV-transition recurrence that derives the partial sum of a target BSV from a reference BSV using sparse add/sub corrections at the flipped positions.

\item We present a Steiner-centered BSV representation and devise a Steiner Absorption Table Construction (SATC) to select a set of Steiner points that maximizes reuse coverage while minimizing the computation. We further selectively prune bit flips to expand the population of reusable BSVs.

\item We design a lightweight ISTS accelerator that reuses Steiner-point partial sums for both static weights and runtime-generated K/V tensors. The hardware integrates lightweight matching and encoder/decoder units to support online BSV matching and compact flip-position storage. Through comprehensive evaluation, we demonstrate that ISTS achieves average speedups of 1.27× and 1.41× over the SoTA Transitive Array and MCBP-BRCR, respectively.
\end{itemize}

\begin{figure}[t]
    \centering
    \includegraphics[width=1\linewidth]{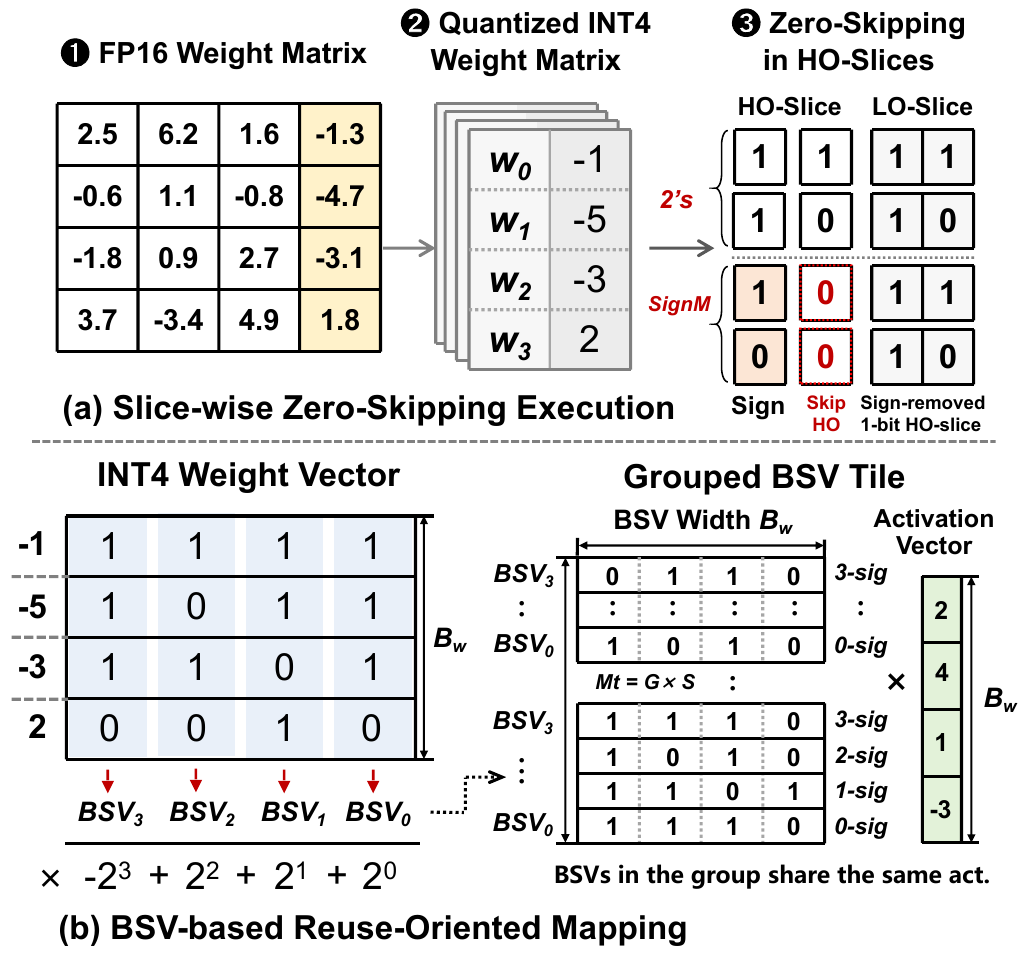}
    \caption{Two bit-slice execution paradigms. (a) Slice-wise zero-skipping execution. (b) BSV-based reuse-oriented mapping.}
    \label{fig2}
\end{figure}

\section{Background and Motivation}
\subsection{Background: Quantization and Bit-slicing}
\label{2.1}

Bit-slice architectures use slices as the basic execution granularity. Existing designs primarily adopt two implementation schemes, as illustrated in Fig.~\ref{fig2}(a) and (b). The first partitions each quantized integer operand into fixed-width slices, such as a 4-bit value into two 2-bit slices, and executes these slices independently~\cite{Bit-slice-Architecture,Lutein,sibia}. Quantized values may be represented in either two's-complement or sign-magnitude form. When all bits in a HO-slice are zero, the corresponding computation can be skipped. The second organization rearranges a group of quantized operands across the bit dimension, collecting bits of the same significance into binary vectors called bit-slice vectors (BSVs)~\cite{MCBP,Transitive-Array}. For an $S$-bit two's-complement weight $w_i$,
\begin{equation}
w_i
=
\sum_{b=0}^{S-2} 2^b W_i^{(b)}
-
2^{S-1} W_i^{(S-1)},
\qquad
W_i^{(b)}\in\{0,1\},
\end{equation}
where $W_i^{(b)}$ denotes the bit at position $b$, while the most significant bit is the sign bit with a weight of $-2^{S-1}$. As shown in Fig.~\ref{fig2}(b), the bits at the same significance across $B_w$ weights constitute one BSV. An INT4 weight vector therefore produces 4 BSVs. Since BSVs derived from weight vectors that share the same activation vector can reuse their partial sums, we collect the BSVs from $G$ such weight matrix into one group. The resulting group contains $M_t=G\times S$ BSVs. For example, $G=64$ and $S=4$ yield 256 BSVs per group. This group defines the reuse domain analyzed in Sec.~\ref{3.2}.

\subsection{Limitations of Existing Bit-Level Acceleration}
\label{2.2}
Table~\ref{tab:reuse_taxonomy} summarizes representative bit-level acceleration mechanisms. Existing approaches can be broadly categorized into three forms: skipping ineffectual bits or slices, reusing repeated values or patterns, and updating prior results through bit-pattern transitions. Prior bit-level sparsity accelerators primarily exploit bit sparsity that is either inherent in the data or induced through pruning\cite{asbp}\cite{bitpruning}\cite{BBS}\cite{Pragmatic}\cite{Bitlet}. Pragmatic~\cite{Pragmatic} and Bitlet~\cite{Bitlet} compute only the partial products associated with nonzero activation and weight bits, respectively. BBS~\cite{BBS} exploits bidirectional bit sparsity to increase skip opportunities and improve workload balance. While these designs skip ineffectual bits to save compute, their execution gains diminish as bit density increases.

Repeated values and patterns enable extensive computation reuse~\cite{UCNN,Panacea,MCBP,PattPIM,RePIM,EACNN}. UCNN~\cite{UCNN} shares convolutional subcomputations across repeated weight values, while CREW~\cite{CREW} memoizes products between each input and its distinct weight values for subsequent indexed reuse. LUT Tensor Core~\cite{LUTTensorCore} optimizes LUT-based LLM inference through table symmetrization and reuse-oriented tiling. However, wider groups exponentially increase LUT size, whereas narrower groups require more lookups and partial-sum accumulations. Panacea~\cite{Panacea} encodes frequent nonzero slices introduced by asymmetric quantization and recovers their contributions through lightweight compensation. MCBP~\cite{MCBP} identifies repeated bit-column patterns and merges their associated activation terms, as shown in Fig.~\ref{fig3}~(b-i). Its reuse requires exact pattern repetition; even a one-bit difference prevents two columns from sharing computation. This limitation becomes more pronounced as the bit width increases. TA~\cite{Transitive-Array} exploits transitive sparsity between binary rows for partial-sum reuse. As shown in Fig.~\ref{fig3}~(b-ii), when the set bits of one row form a subset of another, the latter reuses the former partial sum and accumulates only the activations associated with newly introduced bits. This reuse induces a partial ordering among rows, where long dependency chains in the Hasse graph can lengthen the critical execution path.

\subsection{Opportunities for BSV Reuse}
We identify three challenges in translating BSV reuse into an efficient accelerator and present targeted solutions spanning reuse relations, execution dependencies, and architectural overhead.

\begin{figure}[t]
    \centering
    \includegraphics[width=1\linewidth]{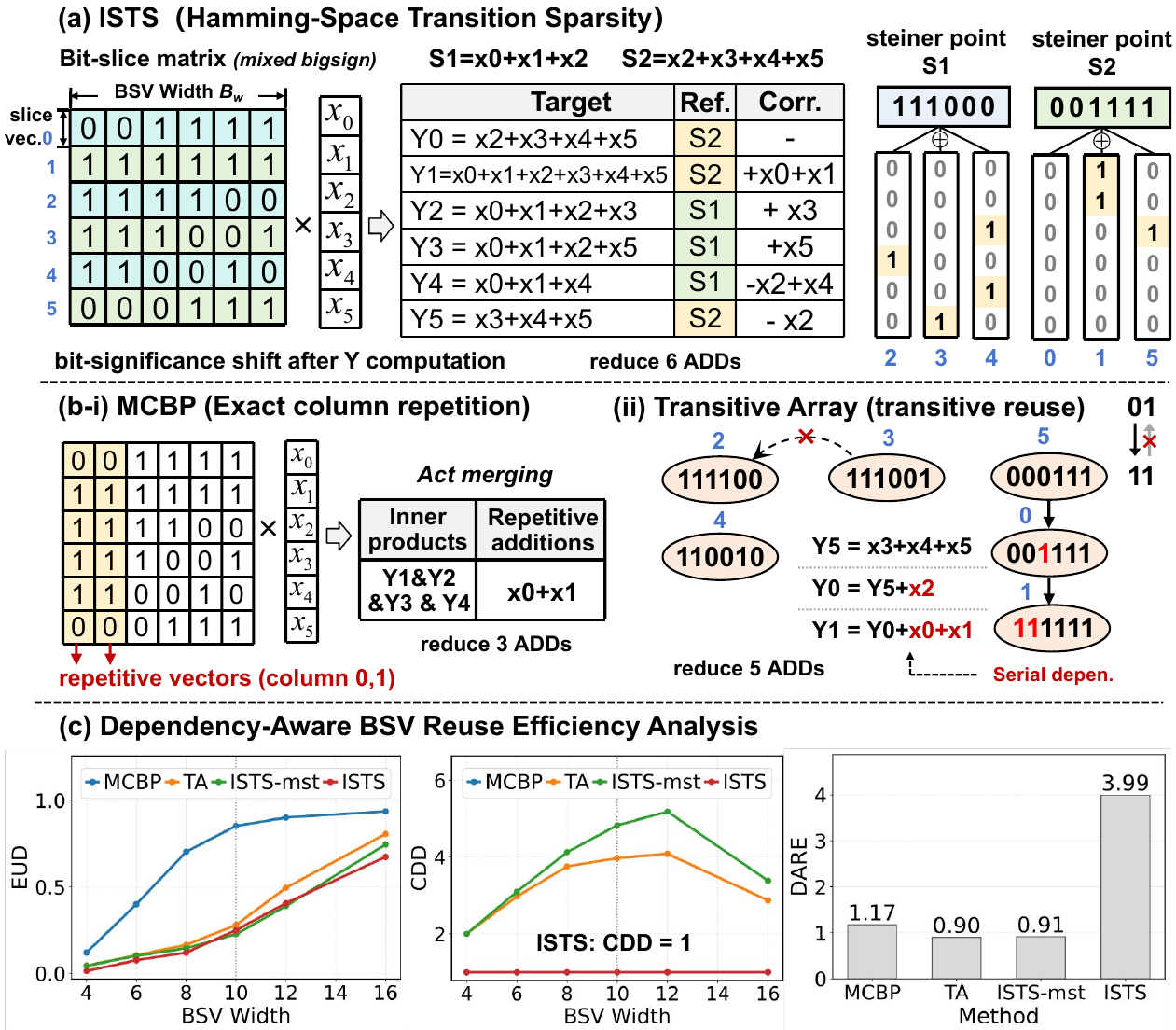}
    \caption{Comparison of ISTS with MCBP and TA. (a) ISTS transition reuse. (b) Reuse mechanisms of MCBP and TA. (c) Effective update density (EUD), critical dependency depth (CDD), and dependency-aware reuse efficiency (DARE) across BSV widths. For a consistent comparison, the MCBP group matrix is transposed so that $B_w$ consistently denotes the length of the reuse pattern. Here, $\mathrm{DARE}=1/(\mathrm{EUD}\times\mathrm{CDD})$. Higher DARE indicates greater reuse efficiency considering computation and dependency depth.}
    \label{fig3}
\end{figure}

\textbf{Challenge 1: The reuse relation determines both reuse scope and computation cost.}
Reusing identical BSVs incurs little update cost but captures limited reuse opportunities, whereas transition-based reuse extends computation sharing to nonidentical BSVs at the cost of additional bit updates. We quantify this trade-off using Effective Update Density (EUD), defined as the average number of effective bit updates per BSV after reuse. As depicted in Fig.~\ref{fig3}~(c), MCBP's EUD approaches 1 as $B_w$ increases because exact repetitions become less frequent within a finite group. In contrast, transition-based schemes reuse computation across nonidentical BSVs and therefore maintain lower EUD.

TA's subset-based transitive reuse cannot capture the reuse between $\mathtt{111100}$ and $\mathtt{111001}$ in Fig.~\ref{fig3}(b-ii), as their transition involves both added and removed 1-bits. More generally, Sperner's theorem~\cite{Sperner1928} shows that distinct BSVs with the same number of 1-bits cannot reuse one another under this relation. For $B_w=10$, the 252 BSVs containing five 1-bits form the largest such group, illustrating how the subset constraint excludes many BSV pairs from reuse. Thus, the choice of reuse relation directly affects both reuse coverage and the computation cost.

\textbf{Key Idea.}
We propose a hamming-space bidirectional BSV transition mechanism that supports both $0\!\rightarrow\!1$ and $1\!\rightarrow\!0$ changes, enabling reuse between arbitrary BSV pairs. The Hamming distance between two BSVs directly determines the required number of bit updates. The formal derivation is presented in Sec.~\ref{3.1}.

\begin{table}[t]
\caption{Comparison of representative bit-level accelerators}
\label{tab:reuse_taxonomy}
\centering
\footnotesize
\setlength{\tabcolsep}{3.5pt}
\renewcommand{\arraystretch}{1.25}

\newcolumntype{R}{>{\hsize=1.2\hsize\centering\arraybackslash}X}
\newcolumntype{N}{>{\hsize=0.8\hsize\centering\arraybackslash}X}
\newcolumntype{D}{>{\hsize=1.0\hsize\centering\arraybackslash}X}

\begin{tabularx}{\columnwidth}{l | R | N | D}
\hline
\textbf{Accelerator} & \multicolumn{1}{c|}{\textbf{Reuse Pattern}} & \multicolumn{1}{c|}{\textbf{Reuse Coverage}} & \multicolumn{1}{c}{\textbf{Dense Support}} \\
\hline

Pragmatic\cite{Pragmatic} & Bit sparsity       & Low       & Limited \\
Bitlet\cite{Bitlet}       & Bit sparsity       & Low       & Limited \\
BBS\cite{BBS}          & 0/1 bit sparsity   & Low       & \checkmark \\
BitWave\cite{BitWave}   & Bit column sparsity & Low       & Limited \\
Panacea\cite{Panacea}   & Frequent slice reuse & Medium    & \checkmark \\
MCBP\cite{MCBP}         & Bit column reuse   & Medium    & \checkmark \\
TA\cite{Transitive-Array} & Transitive reuse   & High      & \checkmark \\
ISTS                   & Bi-transition reuse & High      & \checkmark \\

\hline
\end{tabularx}
\rowcolors{1}{}{}
\end{table}

\textbf{Challenge 2: Reducing computation trades off against execution parallelism.}
Following BSV-pair identification, the resulting reuse relations are organized into an execution topology. By viewing each BSV as a point in Hamming space and each transition as an edge weighted by its Hamming distance, a natural strategy is to construct a minimum spanning tree (MST) that connects nearby BSVs and minimizes the total number of bit updates. However, an MST may yield a deep and unbalanced reuse tree. A deeper BSV must wait for all preceding nodes along its root path to complete, prolonging the lifetime of intermediate partial sums and increasing buffer demand. Variations in tree depth and subtree size may also create uneven workloads across parallel lanes, complicating execution scheduling. For example, TA employs a scoreboard to track dependencies and coordinate node execution. 

Fig.~\ref{fig3}(c) quantifies this trade-off using Effective Update Density (EUD) and Critical Dependency Depth (CDD), where CDD denotes the longest dependency path from a root to a leaf. Both TA and MST-based bidirectional reuse exhibit relatively large CDDs due to their tree-structured dependencies. Compared with TA, the bidirectional MST organization captures more low-flip reuse opportunities and reduces EUD , but introduces deeper dependency trees. 

\textbf{Key Idea.}
We propose a Steiner-based reuse scheme that organizes each BSV group around a small set of Steiner points in Hamming space. Each BSV is assigned to a nearby Steiner point and represented by a sparse bidirectional transition relative to it. Once the Steiner partial sums are available, the associated BSV partial sums can be reconstructed independently in one cycle, limiting the CDD to 1. Fig~\ref{fig3}(a) illustrates this mechanism: six BSVs are assigned to two Steiner points and reconstructed through sparse signed activation corrections.

\textbf{Challenge 3: Expanding reuse coverage increases metadata and hardware overhead.}
Rather than storing the original BSV, ISTS replaces it with a compact descriptor that records the Steiner index and residual bit-flip positions. This overhead becomes particularly important for modern LLMs; for example, DeepSeek-V3 contains 671 billion parameters~\cite{deepseekv3}. Supporting arbitrary-length transitions requires encoding more flip positions, which can make the metadata storage and transfer cost substantially exceed that of the original BSV representation. Moreover, the hardware must provision activation selectors and correction lanes for the maximum transition length, although most BSVs require only a few flips. Therefore, expanding the reuse scope without constraining the transition length may offset the computational savings through increased storage, bandwidth, and hardware overhead.

\textbf{Key Idea.}
To bound metadata and hardware overhead, SATC imposes a limit on transition length. FlipPrune then converts out-of-range transitions into supported forms by pruning low-impact flips, improving reuse coverage without widening the descriptor or correction hardware.

\section{Steiner-Guided BSV Reuse Algorithms}
We develop a three-stage algorithmic framework for efficient BSV reuse. Fig.~\ref{fig7} illustrates the overall flow. We first introduce bidirectional BSV transitions for partial-sum reuse between BSVs (Sec.~\ref{3.1}). Based on this formulation, Steiner Absorption Table Construction (SATC) selects a compact set of Steiner points and assigns nearby BSVs to them, converting recursive BSV-to-BSV reuse into fixed-depth, single-hop updates (Sec.~\ref{3.2}). Finally, FlipPrune expands the reuse range by pruning low-impact bit flips (Sec.~\ref{sec:3.3}).
\subsection{ISTS Theorem}
\label{3.1}

The BSV representation in Sec.~\ref{2.1} reorganizes quantized GEMM into binary-vector computations. This section formulates the ISTS partial-sum reconstruction, which derives a target partial sum from a reusable reference under the same activation vector. It establishes three key properties. First, the reconstruction is lossless, as the target result is obtained by applying add/sub corrections to the reference result. Second, the cost of reuse is determined by the number of changed bit, rather than by the full BSV width. Third, because the recurrence is evaluated before bit-significance shifting, the same transition rule applies to all BSVs.

Consider a reference BSV $\mathbf{v}^{(r)}$ and a target BSV $\mathbf{v}^{(t)}$ that operate on the same activation vector $\mathbf{x}$. Let
$P(\mathbf{v})=\sum_{i=0}^{B_w-1}v_i x_i$
denote the unshifted partial sum of a BSV. Equation~\ref{eq:ists_difference} decomposes the target partial sum into the reference partial sum and a transition term determined by bit differences between the target and reference BSVs.
Because both BSVs are binary, the difference
$v_i^{(t)}-v_i^{(r)}$
directly captures how the contribution of $x_i$ changes: $+1$ for a $0\!\rightarrow\!1$ transition, $-1$ for a $1\!\rightarrow\!0$ transition, and $0$ when the bit remains unchanged. Let $\mathcal{I}_{+}$ and $\mathcal{I}_{-}$ denote the indices of the $0\!\rightarrow\!1$ and $1\!\rightarrow\!0$ transitions, respectively.

\begin{equation}
\begin{aligned}
P(\mathbf{v}^{(t)})
&=
\sum_{i=0}^{B_w-1}v_i^{(t)}x_i \\
&=
\sum_{i=0}^{B_w-1}v_i^{(r)}x_i
+
\sum_{i=0}^{B_w-1}
\left(v_i^{(t)}-v_i^{(r)}\right)x_i \\
\end{aligned}
\label{eq:ists_difference}
\end{equation}

\begin{equation}
P(\mathbf{v}^{(t)})
=
P(\mathbf{v}^{(r)})
+
\sum_{i\in\mathcal{I}_{+}}x_i
-
\sum_{i\in\mathcal{I}_{-}}x_i .
\label{eq:ists_recurrence}
\end{equation}

\begin{figure}[b]
    \centering
    \includegraphics[width=1\linewidth]{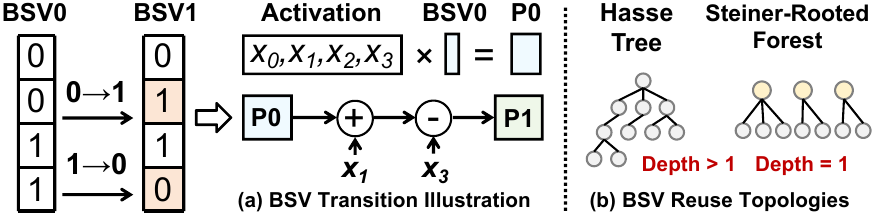}
    \caption{Illustration of BSV transition reuse. (a) Partial-sum update through bidirectional BSV transitions. (b) Comparison of Hasse-tree and Steiner-rooted reuse topologies.}
    \label{fig4}
\end{figure}

ISTS therefore reconstructs a target partial sum by applying activations add/subtracting only at positions where its BSV differs from the reference, rather than recomputing the complete dot product. In the $B_w$-dimensional Hamming space, these positions correspond to the set bits of the XOR mask between the two BSVs, whose population count gives their Hamming distance. The example in Fig.~\ref{fig4}(a) thus requires two correction.

\begin{equation}
r_{\mathrm{tr}}
=
\left\|
\mathbf{v}^{(t)}\oplus\mathbf{v}^{(r)}
\right\|_{0}
=
|\mathcal{I}_{+}|+|\mathcal{I}_{-}|.
\label{eq:transition_rank}
\end{equation}

Each BSV can further be viewed as a point in the $B_w$-dimensional Hamming space, while a reuse relation forms an edge whose weight is determined by the Hamming distance between its endpoints. Selecting different reference points and edges produces different reuse topologies. As shown in Fig.~\ref{fig4}(b), a Hasse tree connects BSVs through multi-level transitions, whereas a Steiner-rooted forest associates each target directly with a shared Steiner point, leading to different dependency depths and execution parallelism.

\subsection{Steiner Absorption Table Construction}
\label{3.2}
Although any BSV can be reconstructed from another through bidirectional bit flips, unconstrained transitions would incur excessive metadata and hardware overhead. To bound these costs, Steiner Absorption Table Construction (SATC) selects $K$ Steiner references for each BSV group and restricts reuse to a hardware-supported flip radius $R$. Each absorbed BSV can therefore be reconstructed with at most $R$ add/sub corrections. SATC jointly selects Steiner references to maximize absorption coverage, while uncovered BSVs are processed through direct computation. It further minimizes the average residual flip count to reduce reconstruction cost. This yields a hardware-bounded coverage problem in binary Hamming space, which is NP-hard in general and reduces to the NP-hard hypercube segmentation problem when $R=B_w$~\cite{feige2014np}. SATC therefore adopts a coverage-driven heuristic for efficient Steiner construction.

Algorithm~\ref{alg:steiner_construction} presents the three-stage process: Step~\ding{182} first selects the seed that covers the largest number of observed BSVs, thereby representing the densest region of the local distribution (Line~2). The remaining seeds are selected by marginal absorption gain, measured as the number of additional BSVs covered by each candidate (Line~5). When candidates provide similar gains, the one farther from the current seed set is preferred to reduce overlap and avoid wasting the limited Steiner counts.
Step~\ding{183} partitions the observed BSVs into $K$ local groups by assigning each BSV to its nearest seed in Hamming space (Lines~8--9). Step~\ding{184} constructs one Steiner point for each group using a bit-wise weighted majority vote over its assigned BSVs. Since the Hamming-distance objective is separable across bit positions, each Steiner bit can be determined independently by the weighted majority at the corresponding position, thereby minimizing the total intra-group distance~\cite{huang1998extensions}. Once the Steiner points are obtained, SATC assigns each observed BSV to its nearest Steiner reference and generates a sparse transition mask that records the differing bit positions (Lines~10-13). The transition mask is then converted into the compact hardware encoding described in Sec.~\ref{4.2}.

\begin{algorithm}[t]
\caption{Absorption-Guided Steiner Construction and Transition Profiling}
\label{alg:steiner_construction}

\KwInput{BSV group $\mathcal{V}$, Steiner count $K$, flip radius $R$}

\KwOutput{Steiner table $\mathcal{S}$, transition profile $\mathcal{P}$}

\SetKwProg{Fn}{def}{:}{}
\Fn{\textsc{ConstructSteiner}$(\mathcal{V}, K, R)$}{
    $\mathbf{s} \leftarrow \textsc{PrimaryAnchor}(\mathcal{V})$\;
    $\mathcal{S} \leftarrow \mathcal{S} \cup \{\mathbf{s}\}$\;

    \Comment{Select seeds for max coverage}
    \While{$|\mathcal{S}| < K$}{
        $\mathbf{s} \leftarrow \textsc{MaxGainSeed}(\mathcal{C}, \mathcal{V}, \mathcal{S}, \mathcal{R})$\;
        $\mathcal{S} \leftarrow \mathcal{S} \cup \{\mathbf{s}\}$\;
    }

    \Comment{Generate Steiner points}
    \For{$iter = 1$ \KwTo $T$}{
        $\mathcal{G} \leftarrow \textsc{AssignToNearestSteiner}(\mathcal{V}, \mathcal{S})$\;
        $\mathcal{S} \leftarrow \textsc{BitwiseMajorityUpdate}(\mathcal{G})$\;
    }

    \Comment{Record flips of each BSV}
    \ForEach{$\mathbf{v} \in \mathcal{V}$}{
        $(id, \mathcal{F}) \leftarrow \textsc{NearestDiff}(\mathbf{v}, \mathcal{S})$\;
        $\mathcal{P} \leftarrow \mathcal{P} \cup \{\langle id, |\mathcal{F}|, \mathcal{F} \rangle\}$\;
    }

    \Return $\mathcal{S}, \mathcal{P}$\;
}
\end{algorithm}

\begin{figure}[t]
    \centering
    \includegraphics[width=1\linewidth]{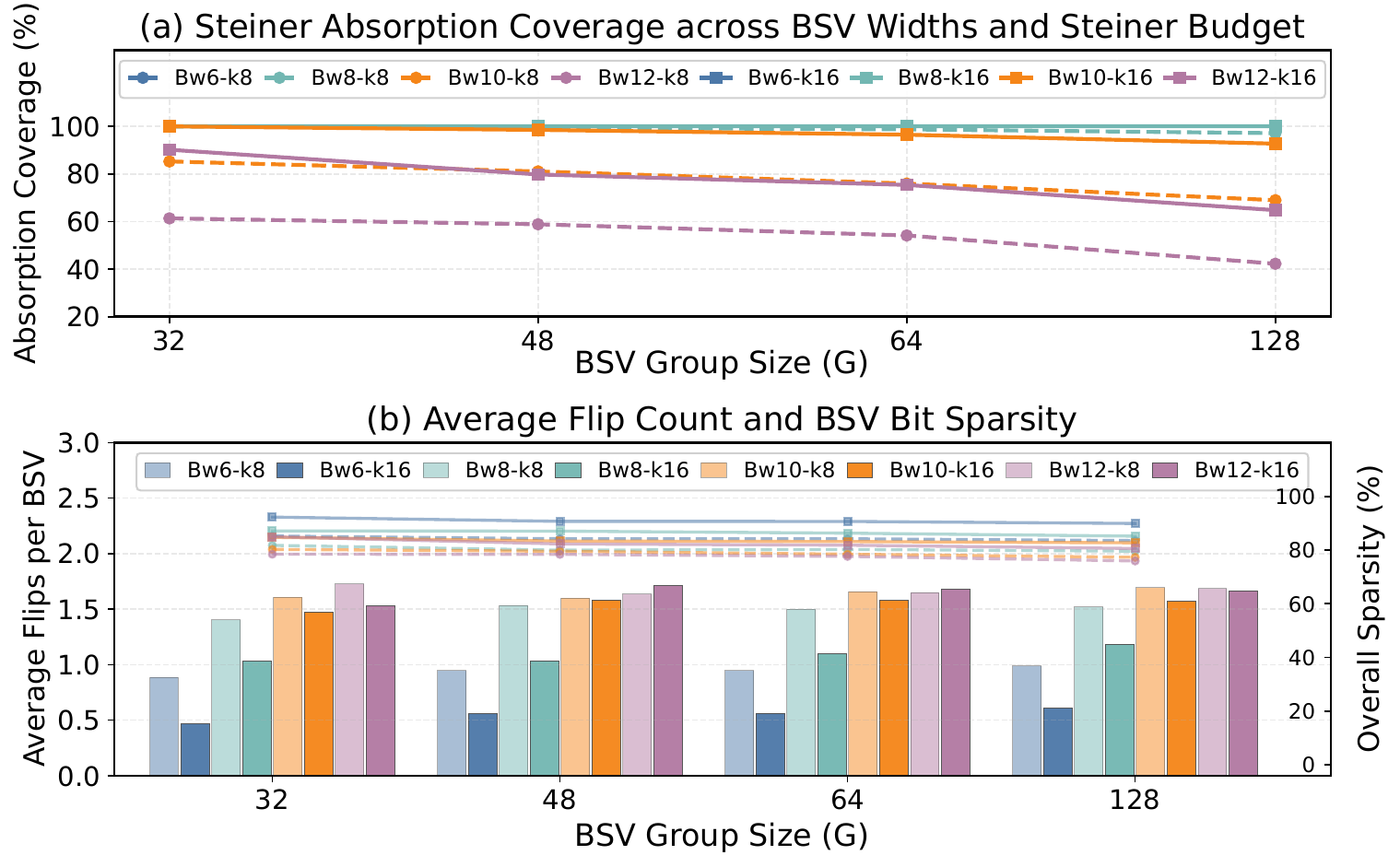}
    \caption{Impact of BSV width and Steiner count on SATC. (a) Absorption coverage under different BSV widths and Steiner counts. (b) Average residual flip count and sparsity.}
    \label{fig5}
\end{figure}

Fig.~\ref{fig5} evaluates the effectiveness of SATC across different BSV widths, group sizes, and numbers of Steiner points. SATC maintains high absorption coverage across the evaluated BSV widths;   even when $B_w$ increases to 10, $K=16$ still absorbs more than 90\% of the BSVs. Fig.~\ref{fig5}(b) further shows that the average transition distance remains only about 1.5-1.7 flips, meaning that more than 80\% of the bits remain unchanged in the transition masks. We compare SATC with three baselines: Top-$K$, which selects the most frequent BSVs; $k$-median, which minimizes the total Hamming distance from each BSV to its nearest reference; and ILP, which solves the problem exactly on small instances. All methods are evaluated on 32 BSV groups sampled from different layers, with 128 BSVs per group and an absorption radius of 2. Top-$K$ and $k$-median achieve coverages of 64.45\% and 79.30\%, respectively. SATC improves the coverage to 93.77\% with an average transition distance of 1.84 flips. The ILP baseline achieves a slightly higher coverage of 95.70\% but requires 12,566.4~s, whereas SATC completes in only 0.16~s, reducing the construction time by $7.85\times10^4\times$. Since ILP does not scale to full-model construction, it is used only as a small-scale reference.

\subsection{Coverage-Driven FlipPrune}
\label{sec:3.3}
Although SATC absorbs about 93\% of BSVs, the remaining BSVs require direct computation and must be stored in RAW format. Each fallback BSV also requires a 6-bit index to identify its position within a 64-BSV block, introducing approximately 3.75\% metadata overhead at $B_w=10$. Since most fallback BSVs are only three flips away from their nearest Steiner reference, we propose \textit{FlipPrune} to remove one low-impact mismatch from eligible BSVs, enabling two-flip reuse while reducing fallback storage and computation.

\begin{figure}[t]
    \centering
    \includegraphics[width=1\linewidth]{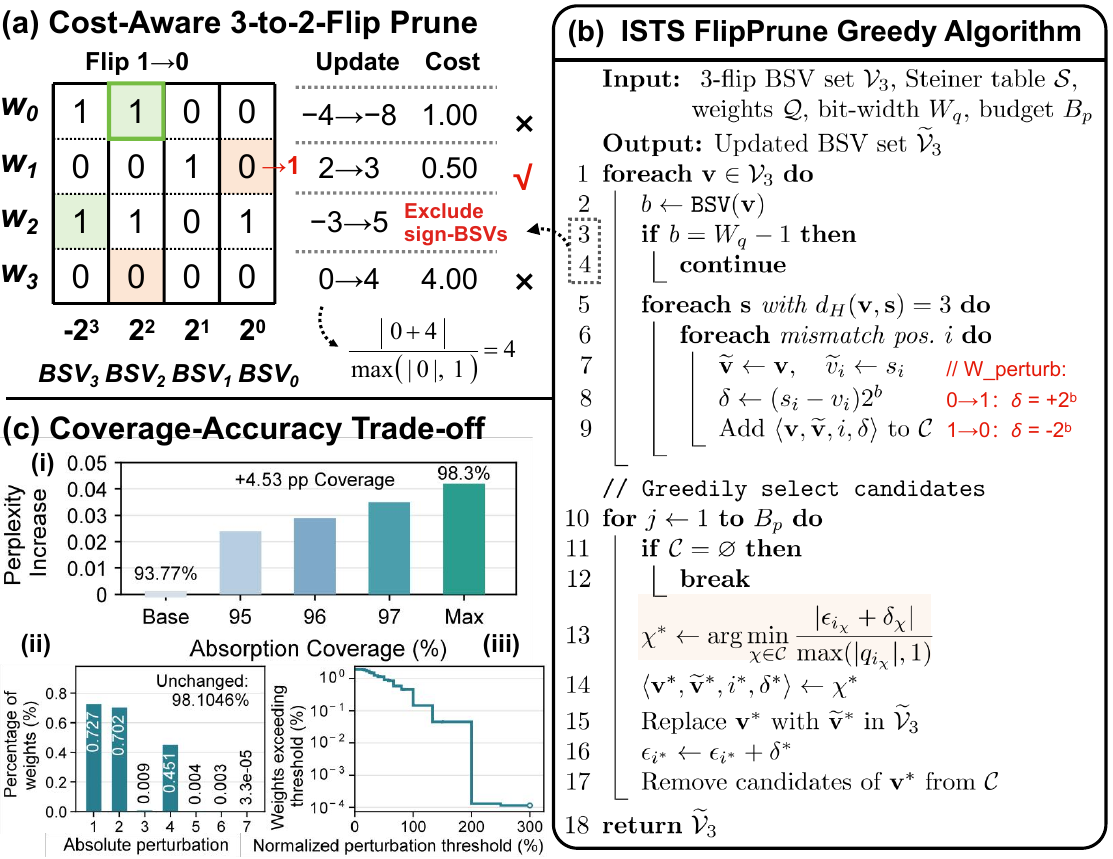}
    \caption{Overview of FlipPrune. (a) Low-impact flips convert 3-flip BSVs into reusable 2-flip forms. (b) A greedy algorithm selects pruning candidates with low weight distortion. (c) Effects of pruning budget on absorption coverage, perplexity, and weight perturbations in LLaMA2-7B~\cite{llama2}.}
    \label{fig6}
\end{figure}

As shown in Fig.~\ref{fig6}(a) and (b), \textit{FlipPrune} consists of three steps. First, Lines~1--10 identify candidate bit modifications. BSVs from the sign bit-plane are excluded to avoid sign reversals and large weight errors. For each remaining BSV $\mathbf{v}$, we examine Steiner references $\mathbf{s}$ satisfying $d_H(\mathbf{v},\mathbf{s})=3$ and identify their mismatched bit positions. Replacing $v_i$ with $s_i$ at any such position reduces the transition distance to two, yielding $\widetilde{\mathbf{v}}$. For bit significance $b$, changing $0\rightarrow1$ increases the corresponding quantized weight by $2^b$, while $1\rightarrow0$ decreases it by $2^b$. The resulting weight change is $\delta_i=(s_i-v_i)2^b$. Second, Lines~11--15 rank the candidates by their normalized cost
\begin{equation}
    c_i=
    \frac{\left|\epsilon_i+\delta_i\right|}
         {\max\left(|q_i|,1\right)},
\end{equation}
where $q_i$ denotes the quantized weight at BSV position $i$. Since \textit{FlipPrune} may also modify other bit of the same weight, $\epsilon_i$ records the weight change introduced by previously selected flips, and $|\epsilon_i+\delta_i|$ evaluates its total change after accepting the current candidate. The denominator is the absolute value of the original weight, so the cost reflects the relative weight change; it is lower-bounded by 1 to handle zero-valued weights. For example, flipping the $2^2$ bit of $w_3$ in Fig.~\ref{fig6}(a) changes its value from $0$ to $4$, resulting in a cost of $4/\max(0,1)=4$, whereas changing $w_1$ from $2$ to $3$ incurs a lower cost of $1/2=0.5$. Finally, Lines~16-18 greedily select and apply up to $B_p$ candidates. After each selection, \textit{FlipPrune} updates the accumulated change of the affected weight and removes all alternatives derived from the same BSV.

As shown in Fig.~\ref{fig6}(c-i), increasing the pruning budget raises absorption coverage from 93.77\% to 98.30\%, while the PPL increases by less than 0.05. The perturbation distributions in Fig.~\ref{fig6}(c)(ii–iii) explain this limited accuracy impact: 98.1048\% of weights remain unchanged, modified weights exhibit mostly small absolute perturbations, and large relative perturbations are rare. Moreover, each transformed BSV modifies only 1 of its 10 bits, resulting in only 0.45\% of all quantized weight bits being altered under the maximum-coverage configuration. Overall, \textit{FlipPrune} substantially improves BSV absorption coverage with negligible accuracy loss.

\subsection{Overall Workflow}
\begin{figure}[t]
    \centering
    \includegraphics[width=1\linewidth]{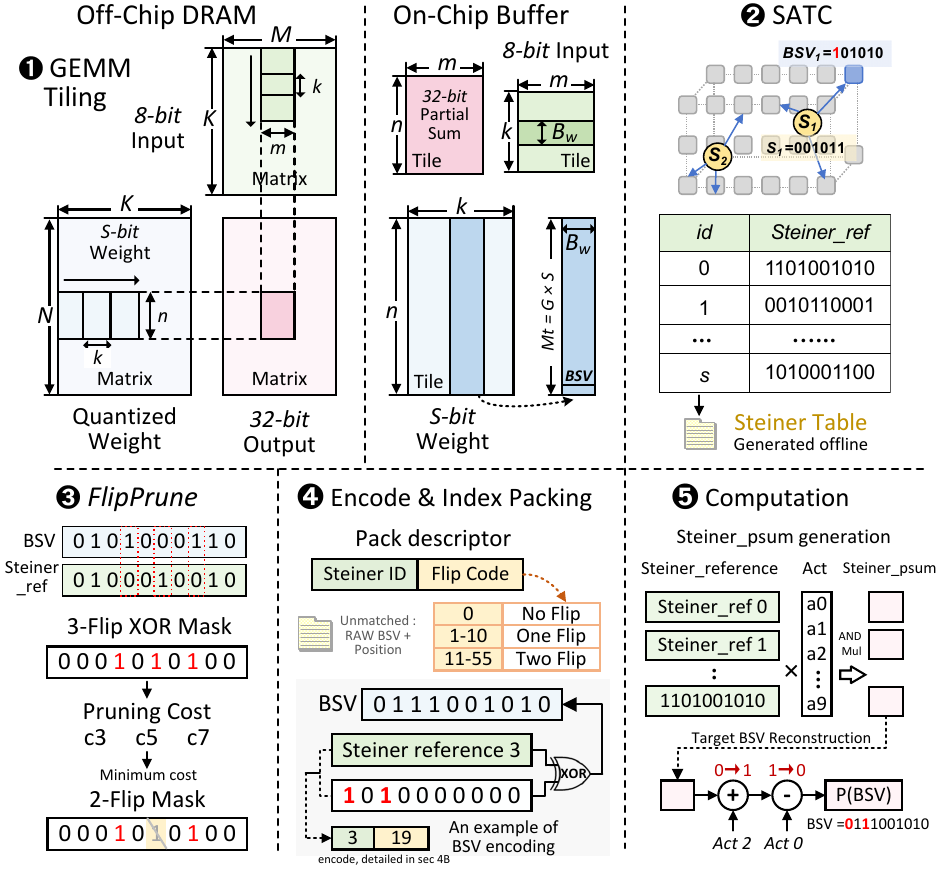}
    \caption{End-to-end workflow of Steiner-based BSV preprocessing, encoding, and hardware execution.}
    \label{fig7}
\end{figure}

Fig.~\ref{fig7} illustrates the end-to-end workflow of ISTS, beginning with GEMM tiling in \ding{182}. Quantized weights and activations are stored in off-chip DRAM and partitioned according to the on-chip buffer capacity and array dimensions. The corresponding tiles are prefetched into on-chip buffers, and each weight tile is further divided into BSV groups that share $K$ Steiner references.
SATC constructs the Steiner references from the BSV distribution of each group in \ding{183}. \textit{FlipPrune} then expands the absorbable BSV population in \ding{184}. Both SATC and \textit{FlipPrune} are performed offline. 
Each absorbed BSV is encoded as a descriptor $\langle\textit{Steiner\_id},\textit{flip\_code}\rangle$ in \ding{185}. For a $B_w$-bit BSV, representing all 0/1/2-flip transitions requires $\binom{B_w}{0}+\binom{B_w}{1}+\binom{B_w}{2}$ codewords. With $B_w=10$ and $K=16$, the 56 transition patterns require a 6-bit \textit{flip\_code}, while the 16 Steiner references require a 4-bit \textit{Steiner\_id}, yielding a 10-bit descriptor with the same width as the original BSV. For example, the target BSV $0111001010$ is encoded as $\langle3,19\rangle$ in Fig.~\ref{fig7} \ding{185}. The \textit{Steiner\_id} selects Steiner reference 3 and its partial sum, while \textit{flip\_code} 19 specifies the two activation positions required for reconstruction. Supporting three flips would increase the descriptor width to 12 bits; thus, BSVs requiring more than two flips retain their original values and position indices for direct computation.
During hardware execution in \ding{186}, the Steiner engine first computes the partial sums of all $K$ references. For each encoded BSV, the \textit{Steiner\_id} selects the reference partial sum, while the \textit{flip\_code} identifies up to two activation positions. The corresponding Steiner bits determine whether each activation is added or subtracted to reconstruct the target partial sum.

\section{ISTS Hardware Architecture}

\subsection{Architecture Overview}
\begin{figure}[t]
    \centering
    \includegraphics[width=0.9\linewidth]{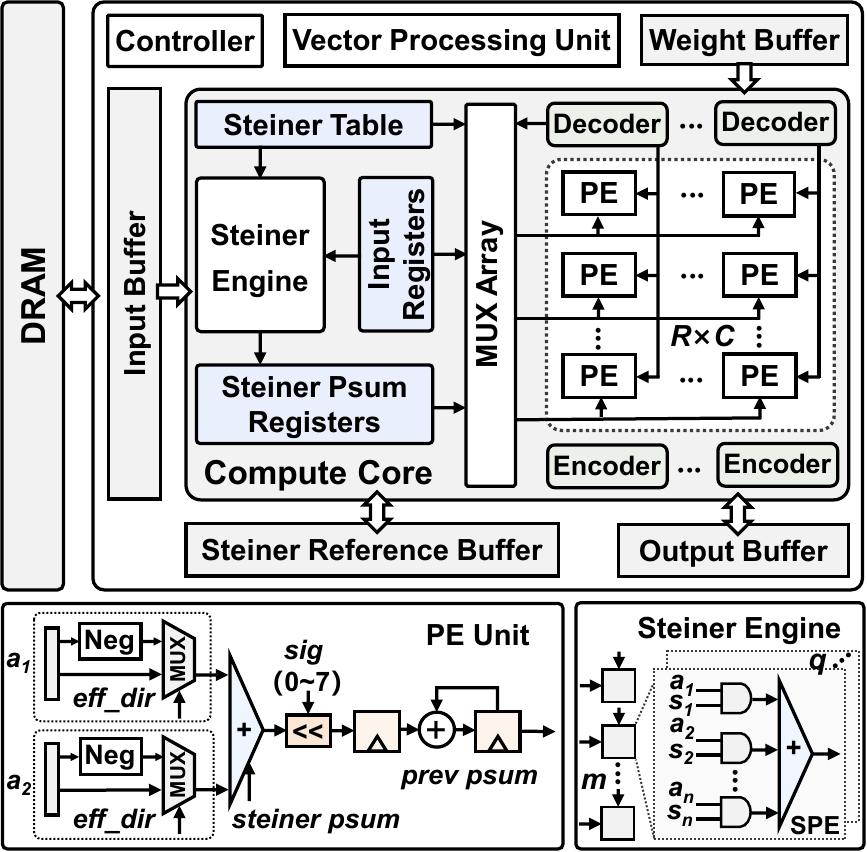}
    \caption{Overview of ISTS architecture.}
    \label{fig8}
\end{figure}

Fig.~\ref{fig8} presents the overall architecture of ISTS, which consists of a top-level controller, a vector processing unit (VPU), four types of L2 buffers, and a compute core. The L2 buffers include input, weight, Steiner reference, and output buffers. The compute core integrates an $R\times C$ PE array, a Steiner Engine (SE), L1 buffers (Steiner Table, Steiner Psum Registers, Input Registers), a MUX array, column-wise decoders and encoders. The VPU supports nonlinear operations, including activation functions, normalization, and softmax.

The ISTS compute core supports Steiner-based computation and encoding for both static weights and dynamic K/V. At the start of each computation round, the Steiner Engine computes Steiner psums from the input and Steiner references using AND-gate arrays and adder trees, with results stored in the Steiner Psum Registers. Based on the decoded metadata, the MUX array selects the required input, Steiner psum, and flip information for each PE, which then reconstructs and accumulates the corresponding psum. For static weights, Steiner tables and BSV encoding are generated offline. For dynamic K/V, Steiner tables are constructed from calibration data offline and preloaded into the compute core. At runtime, each K/V BSV is matched against the Steiner references by the encoder's Steiner absorption unit, and the resulting Steiner ID and sparse transition mask are encoded into the compact representation and written to DRAM for subsequent attention computation.

The ISTS adopts an orthogonal-broadcast, output-stationary dataflow. Input vectors are broadcast row-wise from the Input Registers, while encoded weights are routed through the column decoders and broadcast column-wise from the shared weight buffer. The Steiner references and their psums remain resident within the compute core throughout the processing of the current BSV group. The Steiner Psum Registers adopt double buffering, while the Steiner Table and Input Registers employ triple buffering, enabling concurrent operation of the PE array and SE. When a GEMM tile spans multiple BSV groups along the $K$ dimension, the controller processes these groups sequentially while retaining the output psums in the output buffer. After all $K$-dimension groups have been accumulated, the completed outputs are either forwarded to the VPU for subsequent nonlinear operations or written back to DRAM.

\subsection{Processing Element}
As shown in Fig.~\ref{fig8}, the PE reconstructs the dot-product result of a target BSV from a precomputed Steiner psum. Each PE receives activations, flip information, and a Steiner psum. Two input activation values are routed to two correction lanes. The sign of each correction is determined by the flip direction: a $0\!\rightarrow\!1$ transition adds the selected activation to the Steiner psum, whereas a $1\!\rightarrow\!0$ transition subtracts it. The two signed corrections are then accumulated with Steiner psum. For 0/1-flip patterns, the unused correction lanes are masked to zero. Finally, the reconstructed partial sum is shifted according to \textit{sig} to align with the significance of the current weight bit plane, and is forwarded to the output accumulator for accumulation. Unabsorbed BSVs are dispatched to the Steiner Engine for ordinary vector dot products, following the same execution flow as Steiner psum computation. The PE array and Steiner Engine are jointly provisioned to overlap Steiner psum generation with PE execution while reserving sufficient compute capacity for fallback processing. The controller prioritizes Steiner psum generation and uses the remaining execution slots for fallback requests. 

\subsection{ISTS Decoder and Encoder}
\label{4.2}

The ISTS decoder converts a 6-bit \textit{flip\_code} into the activation indices required for psum correction. A naive implementation uses a 56-entry LUT to support 1 zero-flip pattern, 10 single-flip patterns, and 45 two-flip patterns. Although a single LUT is small, replicating it across all PE columns introduces non-negligible area and routing overhead. To reduce decoder area, we replace the 56-entry LUT with lightweight combinational logic based on a ring-distance mapping. The key idea is to arrange the 10 activation positions on a ring and represent each two-flip pattern using a starting position and the circular distance to the second position. Zero-flip and single-flip patterns are handled by simpler dedicated paths.

\begin{figure}[t]
    \centering
    \includegraphics[width=1\linewidth]{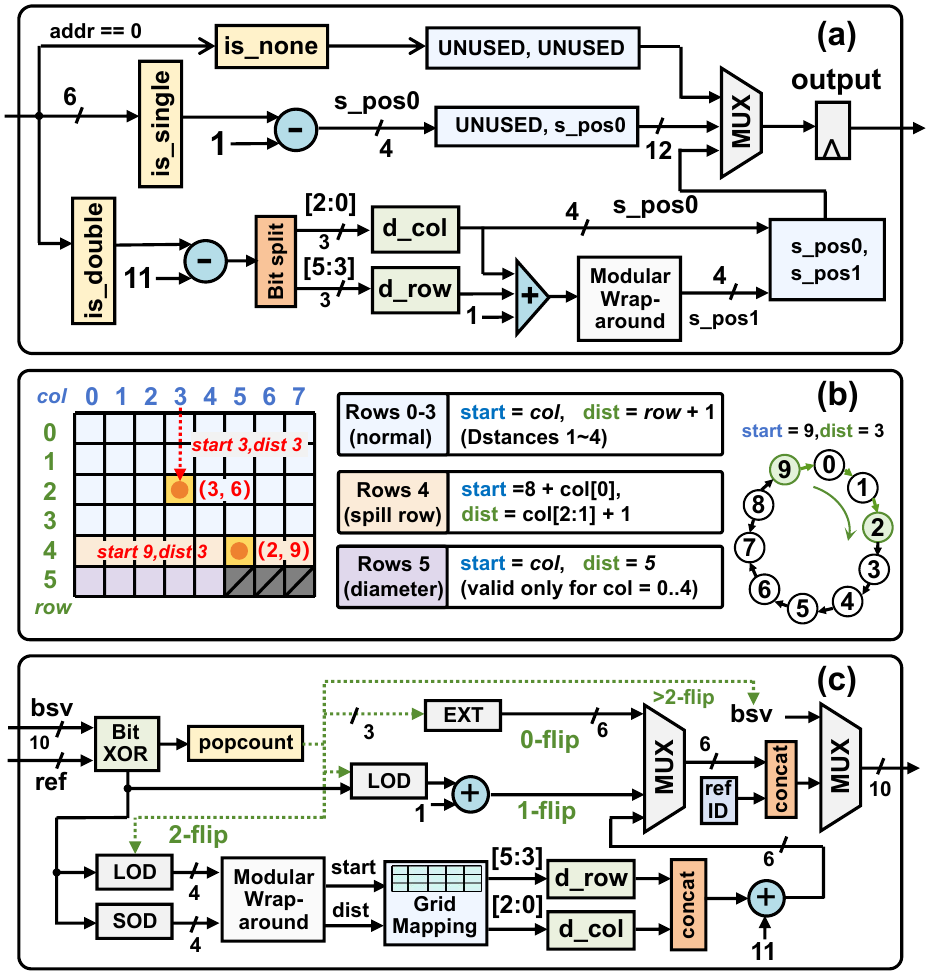}
    \caption{Ring-distance virtual-grid decoding and compact encoding. (a) Decoder. (b) Illustration of the virtual-grid mapping principle for 2-bit flip. (c) Encoder back-end for converting a transition mask into a compact code.}
    \label{fig10_decoder}
\end{figure}

As shown in Fig.~\ref{fig10_decoder} (a) and (b), the ISTS decoder classifies \textit{flip\_code} into 0, 1, or 2-flip patterns using lightweight range comparators. For exact reuse (\textit{flip\_code} = 0), all activation-index outputs are set to the reserved invalid code 15, disabling activation selection. Codes 1–10 directly encode single-flip positions as $\textit{flip\_code}-1$. For two-flip codes (11–55), the decoder computes $k=\textit{flip\_code}-11$ and extracts the coordinates of the $6\times8$ virtual grid through bit slicing: $d_{\mathrm{row}}=k[5:3]$ and $d_{\mathrm{col}}=k[2:0]$. The ring-distance logic then maps these coordinates to the two activation indices.

Fig.~\ref{fig10_decoder} (b) illustrates the interpretation of the virtual-grid coordinate as a ring-distance tuple. For rows 0-3, $d_{\mathrm{col}}$ specifies the first activation index and the starting position on the 10-position ring, while $d_{\mathrm{row}}+1$ gives the ring distance. For example, the grid position at row 3 and column 6 corresponds to a starting position of 6 and a ring distance of 4. The second activation index is therefore computed as $(6+4)\bmod 10=0$. This regular region encompasses start positions from 0 to 7 paired with distances from 1 to 4. To cover the missing start positions 8 and 9 without expanding the grid width, row 4 is designated as a \textit{spill row}, where $d_{\mathrm{col}}[0]$ selects the start position between 8 and 9, and $d_{\mathrm{col}}[2:1]+1$ defines the distance. For example, the grid coordinate $(4,5)$ gives a starting position of $8+1=9$ and a distance of $2+1=3$. The second activation index is therefore $(9+3)\bmod 10=2$, producing the flip pair $(2,9)$.
Row 5 further encodes the distance-5 diameter pairs, where only columns from 0 to 4 are valid and the remaining three cells are left unused. 

For two-flip patterns, bit slicing and row-selection logic decode the virtual-grid coordinates into a starting index and ring distance. An adder with conditional wraparound then computes the second activation index. The decoded 0/1/2-flip results are packed into a unified 12-bit control word: zero flips mark both activation slots as \texttt{UNUSED}, one flip activates one slot, and two flips activate both slots.

As shown in Fig.~\ref{fig10_decoder} (c), the encoder back-end converts the transition mask between a target BSV and its selected Steiner reference into a compact 6-bit \textit{Flip Code}, following the inverse mapping of the decoder. The transition mask is first classified by its flip count: zero- and one-flip cases are encoded directly, while two-flip cases are mapped through the virtual-grid representation. The resulting \textit{Flip Code} is combined with the Steiner ID to form the compact weight representation. Masks with more than two flips bypass the encoding path and retain the original BSV, while the nearest-Steiner matching logic is described in Sec.~\ref{sec4.4}.

\subsection{Steiner Absorption Unit} \label{sec4.4}
The Steiner absorption unit matches each runtime-generated K/V BSV to the nearest Steiner reference within the supported flip radius and generates its sparse transition mask. This operation is performed only during K/V projection, and its cost is amortized across subsequent $QK^{\mathsf T}$ and $SV$ computations. Instead of using a 10-bit population-count unit and a multi-bit minimum-reduction tree for every candidate, we divide the XOR mask into five 2-bit groups and classify each Hamming distance as 0, 1, 2, or greater than 2 using saturating threshold logic. Candidates beyond the supported radius are rejected early, while the nearest valid reference is selected by prioritizing distances of 0, 1, and 2.

\begin{figure}[t]
    \centering
    \includegraphics[width=1\linewidth]{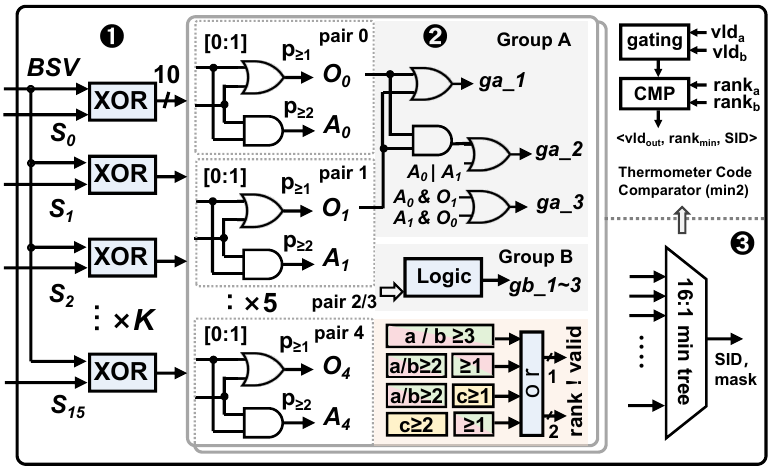}
    \caption{Steiner absorption unit (encoder front-end).}
    \label{fig11_steiner_abs}
\end{figure}

Fig.~\ref{fig11_steiner_abs} illustrates the Steiner absorption unit.
Step~\ding{182} accepts one input BSV and reads the $K$ precomputed Steiner references $\{\mathbf{s}_0,\ldots,\mathbf{s}_{K-1}\}$ from the Steiner table in parallel. For each reference $\mathbf{s}_i$, a $B_w$-bit XOR unit generates the difference mask $\mathbf{m}_i$. In Step~\ding{183}, each 10-bit mask $\mathbf{m}_i$ is divided into five 2-bit subgroups. For subgroup $j$, an OR gate and an AND gate generate $O_{i,j}$ and $A_{i,j}$, respectively. Here, $O_{i,j}=1$ indicates that the subgroup contains at least one set bit, while $A_{i,j}=1$ indicates that it contains two set bits. Subgroups 0 and 1 form Group~A, subgroups 2 and 3 form Group~B, and subgroup~4 is processed separately. A three-level threshold-merging network then combines these local flags. For Group~A and Group~B, the intermediate signals $g_{\ge1}$, $g_{\ge2}$, and $g_{\ge3}$ indicate whether the corresponding 4-bit group contains at least one, two, or three set bits. The final stage combines the threshold signals from Group~A, Group~B, and subgroup~4 to generate $\textit{!vld}_i$ and $\textit{rank}_i$. The signal $\textit{vld}_i$ is deasserted when $\mathbf{m}_i$ contains at least three set bits. For masks within the two-flip radius, $\textit{rank}_i$ encodes Hamming distances 0, 1, and 2 as the 2-bit thermometer codes \texttt{00}, \texttt{01}, and \texttt{11}, respectively. The pair $\langle\textit{vld}_i,\textit{rank}_i\rangle$ is then forwarded to the minimum-reduction tree. \ding{184} The reduction module selects the nearest valid Steiner reference through a hierarchical network of \texttt{min2} units. Invalid inputs are gated off, while two valid inputs are resolved by selecting the smaller thermometer code.

\section{Evaluation}

\subsection{Experimental Setup}
\label{5.1}
\textbf{Baselines.} We compare ISTS against 5 baselines: ANT~\cite{ant}, BBS~\cite{BBS}, LUT Tensor Core~\cite{LUTTensorCore}, TA~\cite{Transitive-Array}, and MCBP~\cite{MCBP}. ANT~\cite{ant} introduces adaptive numerical formats and selects an appropriate format for each tensor. BBS~\cite{BBS} treats either zeros or ones as ineffectual bits and balances the resulting bit-level workloads across processing elements. LUT Tensor Core~\cite{LUTTensorCore} replaces low-bit multiplications with lookup-table to reusable precomputed activation sums, which we extend to a PE array, denoted as LUTA. TA~\cite{Transitive-Array} exploits transitive sparsity by reusing results among bit-sliced rows, with static and dynamic Scoreboard variants denoted as TA-S and TA-D. MCBP~\cite{MCBP} exploits repeated bit-slice columns to reuse computation across recurring column patterns.

\begin{figure*}[t]
    \centering
    \includegraphics[width=1\linewidth]{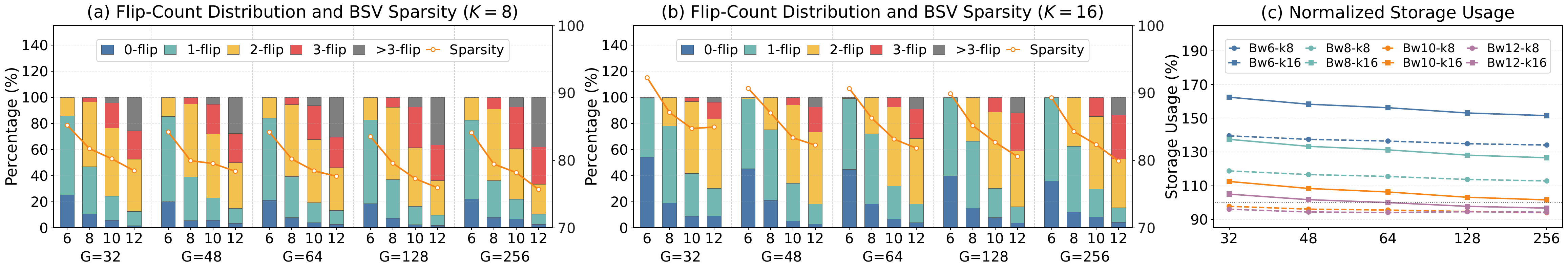}
    \caption{Design-space exploration of ISTS across different BSV widths $B_w$, group sizes $G$, and numbers of Steiner points $K$. (a) and (b) Flip-count distribution and BSV sparsity for $K=8$ and $K=16$. (c) Normalized storage usage.}
    \label{fig12_dse}
\end{figure*}

\textbf{Benchmarks.} We evaluate OPT-1.3B~\cite{opt}, Qwen-2.5-7B~\cite{qwen25}, LLaMA-2-7B, LLaMA-2-13B, and LLaMA-3-8B~\cite{llama2},~\cite{llama3} on the WikiText~\cite{wikitext}, using perplexity (PPL) to evaluate language modeling quality.  We further evaluate downstream accuracy on ARC-Challenge~\cite{clark2018arc}, HellaSwag~\cite{zellers2019hellaswag}, PIQA~\cite{bisk2020piqa}, MMLU~\cite{hendrycks2021mmlu}, and GSM8K~\cite{cobbe2021gsm8k}. Since the performance of several accelerators depends on the weight and activation distributions, we extract layer-wise tensors from the evaluated models as simulator inputs and accumulate the per-layer results to obtain full-model latency and energy.

\textbf{Accelerator Implementation.} We implement the RTL design of ISTS in Verilog, including the compute core, VPU, and global controller, and reproduce the baseline PEs following their original designs. All ISTS and baseline PEs are synthesized using Synopsys Design Compiler with a 28-nm technology library at 500\,MHz for fair comparison. The on-chip SRAM buffers are modeled with CACTI~7.0~\cite{balasubramonian2017cacti} to estimate access latency, bandwidth, and energy. We additionally generate SRAM macros using Memory Compiler and integrate them into the RTL design to validate the area and timing. DRAMSim3~\cite{dramsim3} with a DDR4 configuration is used to model off-chip memory latency and energy consumption. For end-to-end evaluation, we build a cycle-accurate simulator based on ANT~\cite{ant}, retaining its memory-access and energy models while introducing architecture-specific compute models to model the cycle-level scheduling of ISTS, and cross-validate its timing results against RTL simulations. We reproduce the key mechanisms of the other baselines in our simulator, with only MCBP's BRCR modeled for GEMM acceleration.

\begin{table}[t]
    \centering
    \caption{WikiText-2 Perplexity under FlipPrune}
    \label{tab:model_accuracy}
    \footnotesize
    \renewcommand{\arraystretch}{1.2}
    \begin{tabular*}{\columnwidth}{@{} l | @{\extracolsep{\fill}} c ccc c @{}}
        \hline
        \textbf{Model} 
        & \textbf{OPT-1.3B} 
        & \textbf{L-2-7B} 
        & \textbf{L-2-13B} 
        & \textbf{L-3-8B} 
        & \textbf{Qwen2.5-7B} \\
        \hline

        FP16 
        & 13.350 
        & 5.310 & 4.810 & 6.140 
        & 6.850 \\

        INT8 
        & 14.427 
        & 5.483 & 4.897 & 6.244 
        & 6.975 \\

        INT4 
        & 15.440 
        & 5.814 & 5.178 & 7.420 
        & 8.031 \\
        
        \textbf{ISTS (C, 8)} 
        & 14.466 
        & 5.512 & 4.928 & 6.259 
        & 7.015 \\

        \textbf{ISTS (A, 8)} 
        & 14.493 
        & 5.536 & 4.935 & 6.274 
        & 7.031 \\

        \textbf{ISTS (C, 4)} 
        & 15.481 
        & 5.842 & 5.202 & 7.465 
        & 8.065 \\

        \textbf{ISTS (A, 4)} 
        & 15.517 
        & 5.862 & 5.226 & 7.494 
        & 8.087 \\

        \hline
    \end{tabular*}
\end{table}

\begin{table}[t]
    \centering
    \caption{Downstream task accuracy (\%) under W4A8.}
    \label{tab:downstream_accuracy}
    \scriptsize
    \renewcommand{\arraystretch}{1.10}
    \begin{tabular*}{\columnwidth}{@{} l l @{\extracolsep{\fill}} c c c c c @{}}
        \hline
        \textbf{Model} & \textbf{Method} &
        \textbf{ARC-C} & \textbf{Hella.} &
        \textbf{PIQA} & \textbf{MMLU} &
        \textbf{GSM8K} \\
        \hline

        \multirow{3}{*}{LLaMA2-7B}
        & FP16
        & 45.73 & 76.15 & 77.91 & 37.11 & 14.50 \\
        & INT4
        & 43.52 & 73.65 & 76.88 & 32.95 & 12.78 \\
        & ISTS(C,4)
        & 43.47 & 73.59 & 76.80 & 32.89 & 12.72 \\
        & ISTS(A,4)
        & 43.34 & 73.45 & 76.61 & 32.79 & 12.56 \\

        \hline
        
        \multirow{3}{*}{Qwen2.5-7B}
        & FP16
        & 51.11 & 78.94 & 78.78 & 54.53 & 84.50 \\
        & INT4
        & 49.57 & 78.25 & 78.45 & 53.40 & 82.00 \\
        & ISTS(C,4)
        & 49.45 & 78.24 & 78.39 & 53.12 & 85.50 \\
        & ISTS(A,4)
        & 50.17 & 78.16 & 78.35 & 52.96 & 84.50 \\

        \hline
    \end{tabular*}
\end{table}

\subsection{Algorithm Performance}
\label{5.2}
We evaluate the impact of \textit{FlipPrune} on five pretrained language models using perplexity (PPL) on WikiText-2~\cite{wikitext}. All pretrained checkpoints are obtained from Hugging Face~\cite{huggingface} and evaluated using PyTorch~\cite{pytorch}. The INT8/INT4 baselines for OPT and LLaMA2 are generated using OmniQuant~\cite{omniquant}, while LLaMA3.1-8B and Qwen2.5-7B are quantized using the open-source GlowQ framework~\cite{glowq}. We apply \textit{FlipPrune} using conservative (C) and aggressive (A) configurations with target absorption coverages of 97\% and 98.2\%, respectively. As shown in Table~\ref{tab:model_accuracy}, the PPL increase remains below 0.04 and 0.07 for ISTS(C,8) and ISTS(A,8), and below 0.05 and 0.08 for ISTS(C,4) and ISTS(A,4), respectively. Table~\ref{tab:downstream_accuracy} further shows that ISTS(C,4) and ISTS(A,4) maintain downstream accuracy close to the INT4 baselines, with accuracy variations within 3.5\% across all evaluated models and tasks. Minor accuracy gains are within normal evaluation variation. We further analyze FlipPrune-induced modifications across Qwen2.5-7B and LLaMA-2-7B. One-bit modifications affect 1.92\% and 2.05\% of weights, respectively, while two-bit modifications account for only 0.02\% in each model. Three-bit modifications are negligible, at 0.00008\% and 0.00014\%, respectively. Overall, modified bits account for only 0.49\% and 0.52\% of all weight bits, respectively, explaining the negligible accuracy loss introduced by \textit{FlipPrune}.

\subsection{Design Space Exploration}
\label{5.3}

ISTS balances the computational benefits of Steiner-based reuse against storage and hardware overhead through three parameters: BSV width ($B_w$), Steiner count ($K$), and group size ($G$). Increasing $B_w$ reduces storage overhead but shifts more BSVs toward long-distance transitions, while a larger $K$ improves coverage at the cost of matching logic and table storage. Increasing $G$ amortizes each Steiner table over more BSVs, but similarly increases pattern diversity and lowers reuse coverage.

Fig.~\ref{fig12_dse} compares the flip distribution, BSV sparsity, and normalized storage across different $B_w$, $K$, and $G$. At $K=16$ and $G=128$, the fraction of BSVs requiring more than two flips is negligible for $B_w=6$ and $B_w=8$, but increases to $9\%$ and $41\%$ for $B_w=10$ and $B_w=12$, respectively. This degradation is substantially more pronounced with $K=8$, indicating that wider BSVs require more Steiner points to preserve reuse opportunities. Although $B_w=6$ and $B_w=8$ exhibit favorable flip distributions, they incur more than $10\%$ storage overhead, whereas $B_w=12$ reduces storage at the cost of weaker Steiner reuse. Increasing $G$ further amortizes the Steiner-table overhead, but yields only marginal storage savings beyond $G=128$ while continuing to degrade the flip count. We therefore select $B_w=10$, $K=16$, and $G=128$ as the best balance between Steiner reuse and storage overhead.

\begin{table}[t]
    \centering
    \caption{Hardware configuration of ISTS.}
    \label{tab:hardware_config}
    \footnotesize
    \renewcommand{\arraystretch}{1.06}
    \setlength{\tabcolsep}{4pt}
    \setlength{\aboverulesep}{1pt}
    \setlength{\belowrulesep}{1pt}
    \begin{tabularx}{0.92\columnwidth}{
        >{\raggedright\arraybackslash}p{0.31\columnwidth}|
        >{\raggedright\arraybackslash}X
    }
        \toprule
        \textbf{Component} & \textbf{Configuration} \\
        \midrule
        PE Array & $R = 32$, $C = 32$ \\
        \midrule
        Steiner Engine & $m = 32$, $q = 4$ \\
        \midrule
        BSV Width & $B_w = 10$ bits \\
        \midrule
        Steiner References & $K = 16$ \\
        \midrule
        BSV Group Size & $G = 128$ \\
        \midrule
        Input Buffer & 128\,KB, 16 banks, 64-bit/bank \\
        \midrule
        Weight Buffer & 160\,KB, 20 banks, 32-bit/bank \\
        \midrule
        Steiner Buffer & 40\,KB, 10 banks, 32-bit/bank \\
        \midrule
        Output Buffer & 128\,KB, 16 banks, 64-bit/bank \\
        \midrule
        DRAM & DDR4, 2 $\times$ 64-bit channels, 51.2\,GB/s (peak) \\
        \bottomrule
    \end{tabularx}
\end{table}

\subsection{Runtime K/V Analysis}
\label{5.4}
\begin{figure}[t]
    \centering
    \includegraphics[width=1\linewidth]{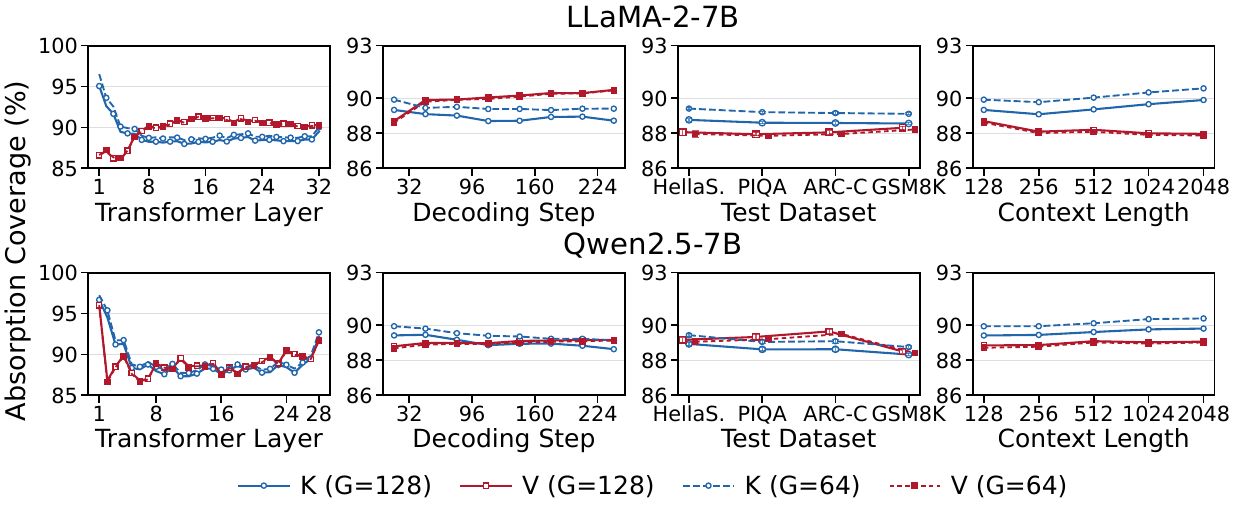}
    \caption{K/V absorption coverage across different workload conditions.}
    \label{fig13}
\end{figure}

ISTS also supports computation reuse for K/V tensors using calibration-derived Steiner tables. Fig.~\ref{fig13} evaluates their absorption coverage across layers, decoding steps, datasets, and context lengths. All experiments use WikiText-2 except the cross-dataset evaluation; prompts contain 128 tokens and generate 32 tokens except in the decoding-step and context-length evaluations. Coverage fluctuates across layers from 86\% to 96\% due to differences in the quantized BSV distributions produced by layer-specific K/V projections, but remains relatively stable across the other dimensions, typically varying by only 1-2\%. Across datasets, Steiner tables capture common BSV patterns shared by different inputs, while the $R=2$ absorption radius allows nearby variants of these patterns to reuse the same references. Coverage also varies little with the group size $G$. Each group size uses a separately constructed Steiner table. For LLaMA-2-7B, 8.943\% of K BSVs can be absorbed only with $G=64$, while 9.039\% can be absorbed only with $G=128$. These nearly equal exclusive fractions offset each other, resulting in similar overall coverage despite the finer-grained $G = 64$. We therefore adopt $G=128$ to reduce Steiner-table overhead. Fallback account for less than 8\% of operations have negligible impact on overall latency. Fig.~\ref{fig16} (c) shows the latency breakdown across hardware modules. Fig. \ref{fig18} separately reports the speedups for linear and attention.  

\begin{figure*}[t]
    \centering
     \includegraphics[width=0.99\linewidth]{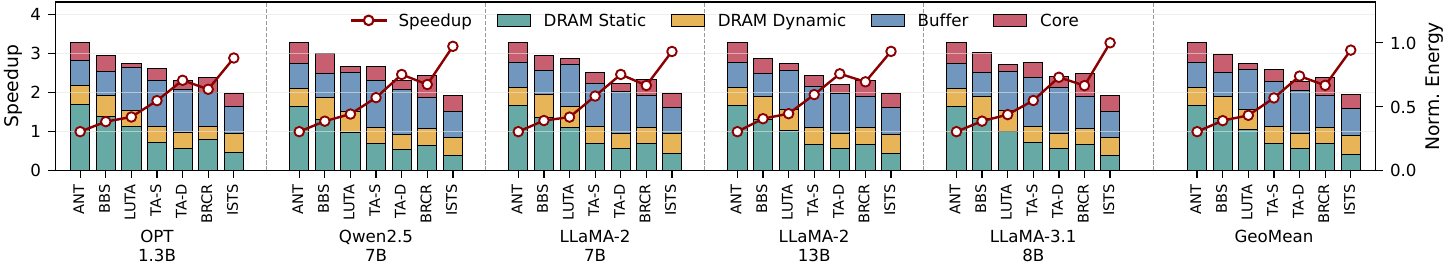}
    \caption{Speedup results and energy breakdown normalized to ANT.}
    \label{fig11}
\end{figure*}

\subsection{Architecture Evaluation}
\label{5.5}
Table~\ref{tab:hardware_config} and Table~\ref{tab:core_area_comparison} presents the configurations and area analysis of ISTS compared to the baseline. ISTS employs a $32\times32$ PE array, and all designs have comparable core areas. Benefiting from the efficient encoding and decoding mechanisms introduced in Sec.~\ref{4.2} and Sec.~\ref{sec4.4}, the proposed decoder and encoder reduce area by 21.8\% and 14.3\% relative to naive implementations of $110.5$ and $688.6\mu\mathrm{m}^{2}$, respectively. They account for only 0.54\% and 3.71\% of the total ISTS core area, respectively, representing modest overheads relative to the PE array. To support bidirectional transition sparsity, each ISTS PE is slightly larger than the combined area of one PPE and one APE in TA. Nevertheless, ISTS requires a smaller buffer than TA. ISTS allocates only a 40\,KB Steiner buffer, whereas TA requires a 90\,KB prefix buffer. This reduction is enabled by the effective reuse of Steiner partial sums generated by the SEs, allowing them to be temporarily held in registers rather than stored in a large L2 buffer.

\begin{table}[t]
\centering
\caption{The area of core components and buffers for ISTS and baseline
architectures using a 28\,nm process.}
\label{tab:core_area_comparison}
\renewcommand{\arraystretch}{1.12}
\setlength{\tabcolsep}{5pt}
\footnotesize

\resizebox{\columnwidth}{!}{%
\begin{tabular}{@{}c|lcc|c@{}}
\toprule
\multirow{2}{*}{\textbf{Architecture}}
& \multicolumn{3}{c|}{\textbf{Core}}
& \multirow{2}{*}{\textbf{Buffer}} \\
\cmidrule(lr){2-4}
& \textbf{Component}
& \textbf{Number}
& \textbf{Area} ($\mathrm{mm}^{2}$)
& \\

\midrule
\multirow{6}{*}{ISTS}
& PE ($261.7\,\mu\mathrm{m}^{2}$)
& 1024
& \multirow{6}{*}{0.510}
& \multirow{6}{*}{456\,KB} \\

& SE ($1708.4\,\mu\mathrm{m}^{2}$)
& 32
& & \\

& Decoder ($86.4\,\mu\mathrm{m}^{2}$)
& 32
& & \\

& Encoder ($590.2\,\mu\mathrm{m}^{2}$)
& 32
& & \\

& MUX Array ($104767.5\,\mu\mathrm{m}^{2}$)
& 1
& & \\

& L1 Buffer ($60495.1\,\mu\mathrm{m}^{2}$)
& 1
& & \\

\midrule
ANT
& PE ($324.8\,\mu\mathrm{m}^{2}$)
& 1536
& 0.499
& 512\,KB \\

\midrule
BBS
& PE ($1170.2\,\mu\mathrm{m}^{2}$)
& 416
& 0.487
& 512\,KB \\

\midrule
\multirow{4}{*}{TA}
& PPE ($53.4\,\mu\mathrm{m}^{2}$)
& 1536
& \multirow{4}{*}{0.536}
& \multirow{4}{*}{480\,KB} \\

& APE ($159.1\,\mu\mathrm{m}^{2}$)
& 1536
& & \\

& NoC ($19520\,\mu\mathrm{m}^{2}$)
& 6
& & \\

& Scoreboard ($92507\,\mu\mathrm{m}^{2}$)
& 1
& & \\

\midrule
\multirow{2}{*}{LUTA}
& PE ($396.9\,\mu\mathrm{m}^{2}$)
& 1280
& \multirow{2}{*}{0.515}
& \multirow{2}{*}{512\,KB} \\

& LUT Entry ($228.56\,\mu\mathrm{m}^{2}$)
&32
& & \\

\midrule
MCBP
& PE ($26417.5\,\mu\mathrm{m}^{2}$)
& 20
& 0.528
& 512\,KB \\

\bottomrule
\end{tabular}%
}
\end{table}

\begin{figure}[t]
    \centering
    \includegraphics[width=1\linewidth]{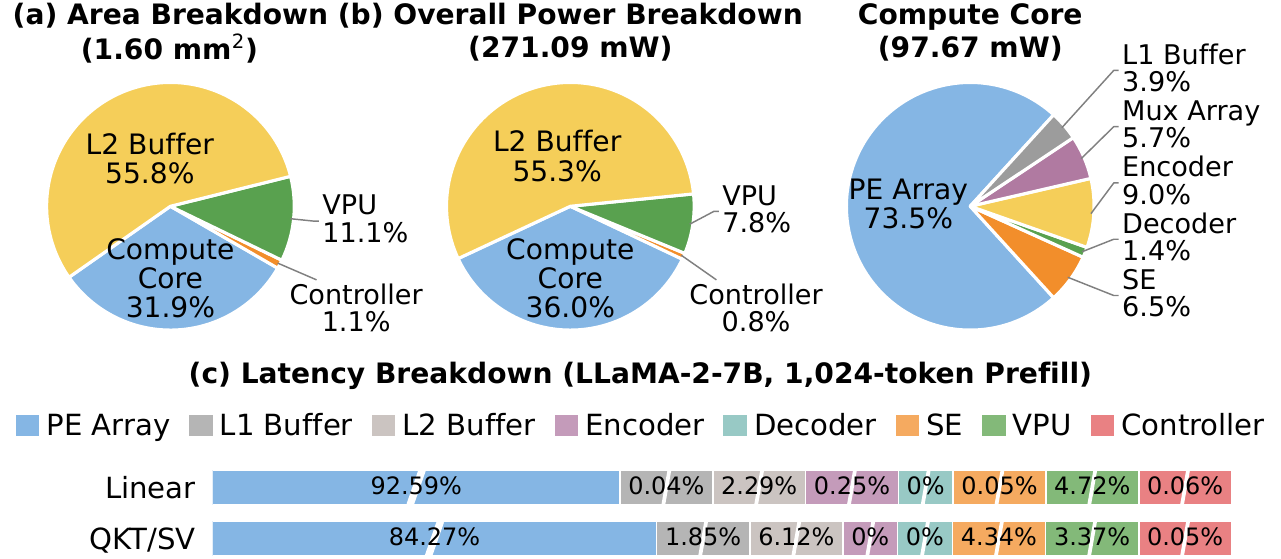}
    \caption{Area, power and latency breakdown of ISTS.}
    \label{fig16}
\end{figure}

Fig.~\ref{fig16} (a-b) presents the area and power breakdown of ISTS, with on-chip buffers dominating both. Within the compute core, the PE array accounts for 73.53\% of power, while the SE consumes only 6.48\% because its partial sums are shared across multiple BSVs. The lightweight decoder further limits its power contribution to 1.36\%, while the encoder accounts for 9.01\% and is activated only for runtime K/V encoding. Fig.~\ref{fig16} (c) shows the module-level latency breakdown, we attribute each cycle to a single module, prioritizing active PE computation and assigning the remaining cycles based on execution dependencies. The PE array dominates latency, while the decoder and encoder contribute negligible exposed latency. The higher fallback rate in $QK^T/SV$ increases the fraction of cycles attributed to the SE. More frequent data reloads and fallback contention also limit overlap between buffer accesses and computation.

\subsection{Comparison with SOTA Accelerators}
\label{5.6}
Fig.~\ref{fig11} shows the speedup and energy breakdown of ISTS and baselines under W4A8 quantization, with a 1024-token prompt, 128-token decoding, and batch size 8. With more effective low-bit quantization, ANT's mixed-precision benefit diminishes, while its adaptive formats increase PE complexity, leading to the lowest performance. BBS performs better by exploiting bidirectional bit sparsity and binary pruning to ensure at least 50\% sparsity. LUTA improves performance through optimized lookup tables but offers limited gains for attention computation, while increasing buffer energy due to LUT storage and frequent table accesses despite reducing computation energy. TA and MCBP further exploit redundancy among BSVs, but MCBP requires exact pattern repetition. TA-D improves performance over TA-S, while its parallelism depends on the distribution of TransRows across Hasse graph. In contrast, ISTS exploits bidirectional transitions while preserving parallel execution, achieving average speedups of $3.09\times$, $2.41\times$, $2.18\times$, $1.27\times$, and $1.41\times$ over ANT, BBS, LUTA, TA-D, and MCBP-BRCR, respectively.
ANT and BBS incur higher core energy due to the hardware overhead of adaptive formats and bidirectional bit skipping. TA incurs high buffer energy due to frequent prefix-buffer accesses, while the matching mechanism in MCBP-BRCR increases compute energy. ISTS instead balances Steiner psum generation, correction, and on-chip buffering, resulting in a more balanced energy distribution between the core and buffers. Across the evaluated models, ISTS improves energy efficiency by $1.67\times$, $1.52\times$, $1.41\times$, $1.17\times$, and $1.22\times$ over ANT, BBS, LUTA, TA-D, and MCBP-BRCR, respectively.

\begin{figure}[t]
    \centering
    \includegraphics[width=1\linewidth]{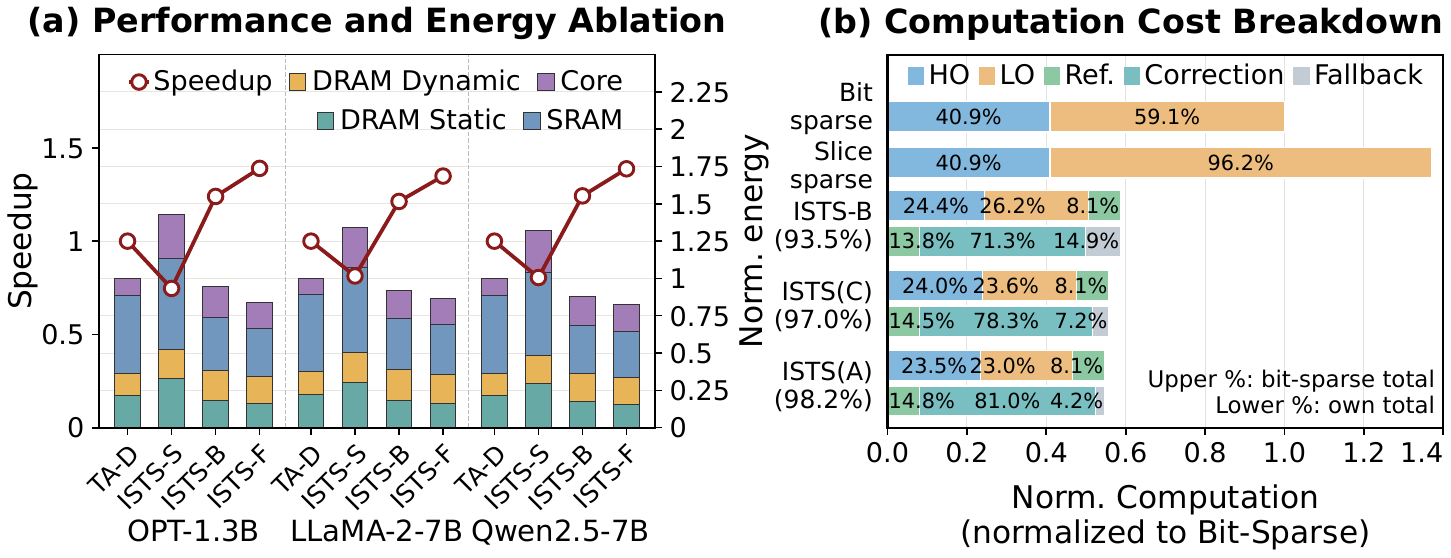}
    \caption{
    (a) Ablation comparing TA-D, ISTS-S, ISTS-B, and Full ISTS with FlipPrune.
    (b) Normalized computation breakdown on LLaMA-2-7B; the upper and lower bars show HO/LO-Slice and reference/correction/fallback computation. }
    \label{fig17}
\end{figure}

\begin{figure}[t]
    \centering
    \includegraphics[width=1\linewidth]{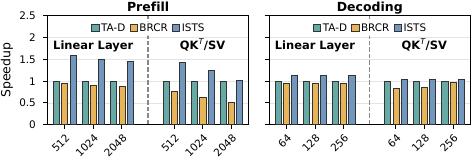}
    \caption{Speedup of linear and attention operators ($QK^T/SV$) on LLaMA-2-7B. Decoding uses a fixed prefill length of 1,024 tokens, batch size = 8.}
    \label{fig18}
\end{figure}

Fig.~\ref{fig17}~(a) evaluates the contributions of the major ISTS mechanisms by progressively comparing TA with a dynamic scoreboard (TA-D), single-direction ISTS (ISTS-S), bidirectional ISTS (ISTS-B), and the full design with FlipPrune (ISTS-F). ISTS-S achieves 69.6\% absorption coverage with unidirectional flipping, leading to frequent fallback and poor performance. Bidirectional transitions increase the coverage to 93.52\%, enabling ISTS-B to achieve a 1.23$\times$ speedup over TA-D. With FlipPrune, ISTS-F further improves performance by 1.12$\times$ over ISTS-B. As shown in Fig.~\ref{fig17}~(b), ISTS-B substantially reduces both HO and LO computation relative to bit-sparse and slice-sparse baselines. Both baselines adopt sign-magnitude representation to increase sparsity. For HO-Slices, ISTS incorporates sign bits into Steiner-based reuse, further reducing computation beyond sparsity-based skipping. FlipPrune further reduces fallback, lowering computation and the storage of fallback position, thereby substantially reducing DRAM dynamic energy.

Fig.~\ref{fig18} compares TA-D, BRCR, and ISTS across varying prefill and decoding lengths, separating linear and $QK^T/SV$ operations. ISTS achieves substantial speedups through bidirectional BSV reuse during compute-intensive prefill, and its gains diminish during memory-bound decoding. During prefill, linear layers achieve higher average speedup (1.51$\times$) than attention (1.22$\times$), benefiting from FlipPrune-enhanced absorption, while dynamic attention incurs more fallback computations. 

\section{Conclusion}

This work presents ISTS, an algorithm-hardware co-designed accelerator that represents each target BSV as a sparse transition from a reusable Steiner point, enabling its partial sum to be reconstructed with only a few additions or subtractions, thereby significantly reducing GEMM computation. Across the evaluated LLM workloads, ISTS achieves speedups of $3.09\times$, $2.41\times$, $2.18\times$, $1.27\times$, and $1.41\times$, and energy-efficiency improvements of $1.67\times$, $1.52\times$, $1.41\times$, $1.17\times$, and $1.22\times$ over ANT, BBS, LUTA, TA-D, and MCBP-BRCR, respectively.

\bibliographystyle{IEEEtranS}
\bibliography{refs}

\end{document}